\documentstyle[prb,aps]{revtex}

\begin{document}
\title{Tilt waves dynamics of the oxygen octahedra in La$_{2}$CuO$_{4}$ from
anelastic and $^{139}$La NQR relaxation}
\author{F. Cordero}
\address{CNR, Area di Ricerca di Tor Vergata, Istituto di Acustica ``O.M. Corbino``,\\
Via del Fosso del Cavaliere 100, I-00133 Roma, and INFM, Italy}
\author{R. Cantelli}
\address{Universit\`{a} di Roma ``La Sapienza``, Dipartimento di Fisica, P.le A.\\
Moro 2, I-00185 Roma, and INFM, Italy}
\author{M. Corti, A. Campana and A. Rigamonti}
\address{INFM-Dipartimento di Fisica ``A. Volta'',\\
Universit\`{a} di Pavia, Pavia, Italy}
\maketitle

\begin{abstract}
The anharmonic vibrational dynamics in nearly stoichiometric La$_{2}$CuO$%
_{4+\delta }$ is studied by means of anelastic and $^{139}$La NQR
relaxation. In the absorption component of the elastic susceptibility as
well as in the nuclear relaxation rate a peak is detected as a function of
temperature, and a relaxation time $\tau =1.7\times 10^{-12}$ $\exp [(2800~$K%
$)/T]~$s is derived. The relaxation processes are attributed to tilt motion
of the CuO$_{6}$ octahedra in doublewell potentials, whose cooperative
character increases the effective energy barrier to the observed value. The
analysis of the relaxation mechanisms has been carried out by reducing the
dynamics of the interacting octahedra to a one-dimensional equation of
motion. The soliton-like solutions correspond to parallel walls separating
domains of different tilt patterns and give rise to pseudo-diffusive modes
which appear as a central component in the spectral density of the motion of
the octahedra. The tilt waves may be considered to correspond to the
dynamical lattice stripes observed in La-based and Bi-based high-$T_{\text{c}%
}$ superconductors.
\end{abstract}

%\draft
\twocolumn

\section{INTRODUCTION}

There is growing interest in the local deviations from the average structure
recently observed in cuprate superconductors. These local inhomogeneities
are often identified with one-dimensional conducting domains of increased
carrier density and antiferromagnetic carrier depleted domains.\cite{BM94}
In some theoretical approaches such stripes are believed to play a role in
the mechanism underlying high-$T_{\text{c}}$ superconductivity. Recently,
the anelastic relaxation spectrum of undoped La$_{2}$CuO$_{4}$ has been
found to exhibit two intense relaxation processes, attributed to intrinsic
relaxational dynamics of unstable lattice modes, namely quasi-rigid tilts of
the O octahedra.\cite{61} The relaxation process below $30~$K is related to
an instability toward the LTT structure, while the high-temperature process
is thermally activated with an activation energy around 2800$~$K. In the
present report combined anelastic and $^{139}$La NQR relaxation measurements
of La$_{2}$CuO$_{4+\delta }$ are presented and interpreted in terms of the
collective dynamics of tilting CuO$_{6}$ octahedra in double-well
potentials. The tilt waves can be considered the equivalent in the undoped
system of the lattice stripes and structural inhomogeneities observed in
high-$T_{\text{c}}$ superconducting oxides.

\section{EXPERIMENTAL AND RESULTS}

The sample was a sintered bar $45\times 4.5\times 0.5$~mm$^{3}$ of La$_{2}$%
CuO$_{4+\delta }$; the excess O was extracted by outgassing at 750~$^{\text{o%
}}$C down to an O$_{2}$ partial pressure of the order of 10$^{-8}$~torr.
Such treatments can introduce few at\% of O vacancies in the CuO$_{2}$
planes,\cite{66} which are reversibly filled after reoxidizing at high
temperature. These vacancies do not affect the anelastic or the NQR
relaxation, being static at the temperatures of interest (below 500$~$K).
The O stoichiometry in the as-prepared state is estimated\cite{66} $\delta
\sim 0.004$. Details on the sample preparation and characterization can be
found in Ref.\ \onlinecite{61}.

The elastic energy loss coefficient $Q^{-1}$, measured by exciting the
flexural vibrations, is given by $S^{\prime \prime }\left( \omega \right)
/S^{\prime }\left( \omega \right) \simeq S^{\prime \prime }\left( \omega
\right) /S\left( \infty \right) $, where $S\left( \omega \right) $ is the
complex dynamic compliance, with unrelaxed value $S\left( \infty \right) $.
The contribution to the imaginary susceptibility from a relaxation process
with characteristic time $\tau $ is \cite{NB} 
\begin{equation}
Q^{-1}\propto [x\left( \delta \lambda \right) ^{2}/T]\,\,\omega \tau
\,/\,[1+\left( \omega \tau \right) ^{2}]  \label{Q-1}
\end{equation}
where $x$ is the atomic fraction of relaxing entities, each producing a
change $\delta \lambda $ of the strain $\varepsilon $ when its state
changes; $\omega /2\pi $ is the measuring frequency. The $Q^{-1}\left(
T\right) $ curve has a maximum at the temperature where $\omega \tau \left(
T\right) =1$, yielding the relaxation rate at the peak temperature. The main
processes appearing in the anelastic spectrum of La$_{2}$CuO$_{4+\delta }$
above 100$~$K are due to the diffusive jumps of the interstitial O atoms.
The peak attributed to the tilts of the O octahedra appears only after
reducing the content of excess O down to very low levels ($\delta <0.001$).
In fact the interstitial O atoms block the relaxational motion of several
surrounding octahedra.\cite{61} The anelastic peak in the nearly
stoichiometric state is shown in Fig. 1 (left hand scale). The two sets of
data refer to the excitation of two flexural modes (460~Hz and 5.9~kHz). The
peak due to O diffusion was 30 times smaller than in the as prepared state,
indicating a concentration of residual oxygen $\delta \sim 1\times 10^{-4}$;
such a small contribution has been subtracted in Fig. 1.

The $^{139}$La NQR relaxation rate $W_{\text{Q}}$ in the same sample is also
shown in Fig. 1 (right hand scale), again after subtraction of the
contribution from the residual interstitial O which gives a monotonous
increase of $W_{\text{Q}}$ for $T\ge 400~$K.\cite{RBL} In La$_{2-x}$Sr$_{x}$%
CuO$_{4+\delta }$ the $^{139}$La NQR relaxation can be due to magnetic or
quadrupolar mechanisms.\cite{BCR89,CRT95} The magnetic relaxation is due to
the fluctuation of the field at the La site, arising from the time
dependence due to the Cu$^{++}$ spin operators. The quadrupolar mechanism is
due to the fluctuation of the transverse components of the electric field
gradient at the La site, mainly caused by the vibrations of the surrounding
O ions, or by a diffusing interstitial O atom passing close to the La
nucleus. The two types of relaxation are described by different recovery
laws,\cite{Reg91} so allowing the identification of the relaxation mechanism
by irradiating different NQR lines. The coincidence in Fig. 2 of the
recovery plots of the NQR echo signals at $2\nu _{\text{Q}}$ and $3\nu _{%
\text{Q}}$ ($\nu _{\text{Q}}=3eV_{zz}\,Q/[2I(2I-1)]$ $\approx $ 6.2~MHz, $eQ$
is the $^{139}$La quadrupole moment) for $T>77~$K indicate that above this
temperature the mechanism is totally quadrupolar, i.e. due to the diffusion
of interstitial O (subtracted in Fig. 1) and to the motion of the apical O
atoms. The relaxation rate $\tau ^{-1}$, determined from the condition $%
\omega \tau =1$ at the maxima of the $Q^{-1}\left( T\right) $ and $W_{\text{Q%
}}\left( T\right) $ curves, is plotted in Fig. 3 in logarithmic scale
against the reciprocal of temperature. The three points can be closely
fitted by a straight line $\tau ^{-1}=\tau _{0}^{-1}\exp \left( -E_{\text{eff%
}}/k_{\text{B}}T\right) $, $\tau _{0}=1.9\times 10^{-12}$~s and $E_{\text{eff%
}}/k_{\text{B}}=2800~$K, clearly indicating that the same process is
observed by both techniques.

\section{DISCUSSION}

The relaxation character of the responses imply that the apical O atoms in
the LaO planes perform pseudo-diffusive motion in a strongly anharmonic
potential. Such a potential should be associated with the softer lattice
modes of the rigid-unit type, which are low frequency modes involving
relative rotations of rigid lattice units, with little or no distortion of
the units (which would increase the restoring force and frequency of the
mode). One can assume that the displacement of an apical O atom represents
the rotation of the rigid octahedron, and both the anelastic and NQR
relaxation are due to the switching of the octahedra between different tilts
in multiwell potentials. Such potentials have been calculated for La$_{2}$CuO%
$_{4}$\cite{CPK} and optimally doped La$_{2-x}$Ba$_{x}$CuO$_{4+\delta }$\cite
{PCK} and La$_{2-x}$Sr$_{x}$CuO$_{4+\delta }$.\cite{BMF} Minima were found
for tilts corresponding to displacements of the apical O atoms by $\delta
\sim \pm 0.2$~\AA\ in the $[110]$ directions (parallel to the in-plane CuO
bonds), corresponding to the low-temperature orthorhombic (LTO) structure,
separated by an energy barrier of $E=35-53$~meV ($400-600~$K). Additional
minima are found in correspondence to the low-temperature tetragonal (LTT)
structure with tilts about the $\left[ 100\right] $ directions,\cite{CPK}
but the LTT domains can be observed by diffraction experiments only when
doping around $\frac{1}{8}$ by Ba or combined Sr and Nd substitution.\cite
{MWZ98} In the following both the anelastic and NQR measurements are
interpreted in terms of cooperative motion of the octahedra in such
multiwell potentials.

Let us first consider the $^{139}$La NQR relaxation due to the atomic
displacements $s\left( t\right) $ of the apical O atoms surrounding the La
nucleus. In order to simplify the interpretation, we will assume
uncorrelated motion of such atoms, and approximate the relaxation rate $W_{%
\text{Q}}$ in the form\cite{Rig84} 
\begin{eqnarray}
W_{\text{Q}}\left( \omega \right) &\simeq &a\left( \frac{eQ}{\hbar }\right)
^{2}\left( \frac{\partial V_{zz}}{\partial s}\right) ^{2}\int \left\langle
s\left( t\right) s\left( 0\right) \right\rangle e^{-i\omega t}dt=
\label{W_Q} \\
&=&A^{2}\cdot J\left( \omega \right)  \nonumber
\end{eqnarray}
where $a$ is a constant of the order of $5/8I\simeq 0.178$, $\left\langle
s(t)s(0)\right\rangle $ is the autocorrelation function of the motion of the
apical oxygen and $J\left( \omega \right) $ the corresponding spectral
density. If $s\left( t\right) $ can switch between two values $\pm \delta $
corresponding to two tilt orientations with probability $1/\left( 2\tau
\right) $, then the correlation function is $\left\langle
s(t)s(0)\right\rangle =\delta ^{2}e^{-\left| t\right| /\tau }$ and the NQR
rate becomes 
\begin{equation}
W_{\text{Q}}=A^{2}\delta ^{2}\,2\tau \,/\,[1+\left( \omega \tau \right)
^{2}]\,.  \label{NQRrelax}
\end{equation}
where $\tau $ is the correlation time.

The imaginary part of the elastic compliance $S^{\prime \prime }\left(
\omega \right) $ can be expressed in terms of the spectral density of the
autocorrelation function of the macroscopic strain $\varepsilon \left(
t\right) $\cite{WK96} 
\begin{equation}
S^{\prime \prime }\left( \omega \right) =(\omega V\,/\,2k_{\text{B}}T)\,\int
\left\langle \varepsilon \left( t\right) \varepsilon \left( 0\right)
\right\rangle e^{-i\omega t}dt\,,  \label{S''}
\end{equation}
$V$ being the sample volume. In the assumption that $\varepsilon \left(
t\right) $ is directly related to the displacements $s\left( t\right) $ of
the O apical atoms, Eqs.~(\ref{Q-1}) and (\ref{S''}) differ from Eqs.~(\ref
{W_Q}) and (\ref{NQRrelax}) only by a factor $\omega /T$ and by the
constants expressing the dependence of strain and NQR frequency on the
atomic displacements.

Now it must be explained why the apparent activation energy $E_{\text{eff}}$
for the switching rate $\tau ^{-1}$ is higher than the theoretically
estimated local potential barrier $E$.\cite{CPK,PCK,BMF} This is related to
the cooperative character of the motion of the octahedra, which has been
widely discussed by Markiewicz.\cite{Mar93b} He considered an LTT ground
state and observed that, neglecting intraplane interactions, the octahedra
along directions perpendicular to the tilt axes are strongly correlated,
since they share the O atoms which are displaced from the CuO plane, while
adjacent rows are weakly correlated, since the shared O\ atoms remain in the
CuO plane. The system is thus reduced to a one-dimensional array of rows of
octahedra, where all the octahedra belonging to the same row are tilted
according to the same pattern. The resulting equations of motion are
non-linear and admit solitonic solutions, which correspond to
one-dimensional propagating walls between domains with different tilt
patterns.\cite{Mar93b} This analysis allows one to apply the one-dimensional
models of non-linear lattice dynamics which have been used to describe
structural phase-transformations,\cite{BMF} as well as the correlated
dynamics of off-centre atoms in perovskites.\cite{TR87} One can consider a
potential of the form $V=\sum_{i}-as_{i}^{2}+\,bs_{i}^{4}+\,\overline{c}%
s_{i}s_{i+1}$, where each atom moves in two minima separated by an energy
barrier $E=(a^{2}/4b)$ with a bilinear coupling to the neighbors. The
coupling constant $\overline{c}$ takes into account a cluster average over
configurations. The resulting equation of motion has solitonic solutions
similar to those found for the octahedra in La$_{2}$CuO$_{4}$,\cite{Mar93b}
and it can be proved\cite{TR87} that the spectral density $J\left( \omega
\right) $ contains both a resonant peak at the frequency of vibration in
each well, and a central peak of pseudo-diffusional character with
characteristic frequency $\tau ^{-1}=\tau _{0}^{-1}\exp \left( E_{\text{eff}%
}/T\right) $ with $E_{\text{eff}}\approx 1.75E\sqrt{2\overline{c}/a}$, $\tau
_{0}=\frac{d}{v}\sqrt{\overline{c}/a}$, where $v$ is the average velocity of
propagation of the soliton-like excitation through the atoms spaced by $d$.
These solitonic solutions provide the observed relaxational contribution to
the spectral density and susceptibility. The effective potential barrier $E_{%
\text{eff}}$ is increased with respect to the local barrier $E$ by the
interaction between the octahedra.

In fitting the experimental data with the above model, a distribution of the
values of $\bar{c}$ has been introduced, corresponding to the distribution
in size and shape of the regions where the octahedra clusters build up the
cooperative dynamics. Such regions are limited by the excess O atoms, which
block the relaxational dynamics. The interaction parameter $\overline{c}$
was distributed according to a gaussian, resulting in a distribution of
effective energy barriers $E_{\text{eff}}$.

A feature of the anelastic peak which cannot directly be accounted for by
the above formulas is the increase of its intensity at higher temperature,
instead of a decrease as expected from Eq. (\ref{Q-1}). Such a temperature
dependence is observed in the case of relaxation among states which differ
in energy by $\Delta E\geq k_{\text{B}}T$. In this case the relaxation
strength must be multiplied by a factor containing the product of their
equilibrium occupation numbers,\cite{35} $4n_{1}n_{2}=\left[ \cosh \left(
\Delta E/2k_{\text{B}}T\right) \right] ^{-2}$. This causes a decrease of the
relaxation strength below $T\sim \Delta E/k_{\text{B}}$. The relaxation rate
must be multiplied by $\cosh \left( \Delta E/2k_{\text{B}}T\right) $, but
this correction does not affect sizably the relaxation curves. The
expression of the NQR rate should be modified in the same manner.\cite{LL66}
It should be remarked that these corrections are strictly valid for
relaxation between levels without cooperative effects, and their extension
to the above model of cluster dynamics is not obvious.

The resulting fit is shown in Fig. 1 as continuous lines. The mean values $%
\tau _{0}=$ $1.7\times 10^{-12}~$s and $E_{\text{eff}}=2800~$K are the same
as deduced from the condition $\omega \tau =1$ at the maxima (Fig. 2). The
effective barrier $E_{\text{eff}}$ is compatible with the theoretically
estimated\cite{CPK,PCK,BMF} local barrier of $\symbol{126}500~$K with a mean
coupling constant $\overline{c}_{0}\simeq 5a$. The width of the distribution
of $\bar{c}$ is $\sigma =0.18\overline{c}_{0}$ for the NQR and $0.25%
\overline{c}_{0}$ for the anelastic\ data (a temperature dependent width may
result from the ordering of interstitial O); the resulting distribution in $%
E_{\text{eff}}$ is gaussian with a width of $\sim 300~$K. The asymmetry
energy $\Delta E=280~$K is 10 times smaller than $E_{\text{eff}}$ and
therefore it does not affect the overall picture.

The fit value for the strength of the NQR relaxation mechanism is $A\cdot
\delta =74$~kHz; $A$ can be estimated in the hypothesis of independent
displacements of the O atoms, assuming $\partial V_{zz}/\partial s\simeq $ $%
3V_{zz}/d$ ($d\simeq 2.7$~{\AA } is the La-O distance), and for $\delta \sim
0.2$~\AA\ it yields a relaxation strength $A\cdot \delta \simeq $ $\left(
3\omega _{\text{Q}}/d\right) \delta \simeq $ $8.6$~MHz, about a factor 100
larger than the experimental one. A reduction of the strength of the
relaxation mechanism may be due to the correlation between the displacements
of the five apical O atoms surrounding the La nucleus, giving rise to a form
factor which multiplies the spectral density. This correction, however, can
hardly produce a reduction by 100 times, which is likely to reflect the
fraction of octahedra which are instantaneously changing the tilt
orientation. In a simple picture one can consider independent solitonic
domain walls (DW), corresponding to tilt waves which propagate through the
lattice with velocity $v$; each La nucleus contributes to the NQR relaxation
only when run over by the tilt wave, so that the relaxation strength should
be reduced by the fraction of octahedra within the domain walls.

Along the same line, one could attribute the observed increase of the
anelastic relaxation intensity with temperature to a temperature dependent
volume fraction of DW, instead of introducing the asymmetry energy $\Delta E$%
. An increase of DW density, roughly as $\left( T_{t}-T\right) ^{-1}$, is
indeed observed on approaching martensitic transformations,\cite{HWS92} but
is not sufficient to explain the increase of the anelastic relaxation
strength by 30\% from 147 to 170$~$K (Fig. 1, taking into account the $1/T$
factor in Eq. (\ref{Q-1})). In the present case, the DW exist below the
transformation from the high temperature tetragonal structure to the LTO
structure at $T_{t}=530~$K and their density would increase of only 6\% on
increasing the temperature from 147 to 170$~$K.

From the close coincidence of the relaxation rate describing the anelastic
and NQR results, it is deduced that the propagation of soliton-like walls is
the origin of both types of relaxation, and that there is a direct
relationship between the atomic displacement $s$ and macroscopic strain $%
\varepsilon $ whose spectral densities are probed by the two experiments.
Such a direct relationship can be assumed for the movement of a wall
separating two LTO domains with opposite signs of the in-plane shear strain $%
\varepsilon _{xx}-\varepsilon _{yy}$ (in LTO notation, with $a$ and $b$
directions at $45^{\text{o}}$ with the CuO bonds). The propagation of such
DW changes the domain sizes and therefore modulates $\varepsilon $.

The spatial arrangement of the domains of tilted octahedra cannot be deduced
by our data, but the above analysis corresponds to an array of propagating
parallel fronts of tilt waves. Similarly, the picture proposed by Markiewicz%
\cite{Mar93b} for La$_{2-x}$Sr$_{x}$CuO$_{4+\delta }$ is a LTT ground state
with lowest energy excitations consisting of parallel propagating LTO walls.
A closely spaced structure of alternating LTT domains and propagating LTO
walls would form a so-called dynamic Jahn-Teller phase with average
orthorhombic structure, and conventional diffraction measurements would see
the average structure (the intermediate structure between two LTO domains is
LTT and viceversa). Such a microstructure closely corresponds to the lattice
stripes proposed by Bianconi {\it et al}. on the basis of EXAFS results in
the same doped material La$_{1.85}$Sr$_{0.15}$CuO$_{4+\delta }$\cite{BSL}
and in Bi$_{2}$Sr$_{2}$CaCu$_{2}$O$_{8+y}$.\cite{BM94} Stripes have also
been observed by neutron diffraction\cite{TSA} in La$_{1.6-x}$Nd$_{0.4}$Sr$%
_{x}$CuO$_{4}$ with $x=0.12$, where they are pinned by the formation of a
stable LTT structure. Also our $^{139}$La spectra exhibit asymmetric
broadening of the lines split by the local magnetic field observed by other
authors,\cite{RHA} supporting the picture of channels where the charges
doped by the residual excess O are mobile and depress the antiferromagnetic
order.

\section{CONCLUSION}

In summary, thermally activated relaxation processes due to strongly
anharmonic lattice modes have been evidenced in nearly stoichiometric La$%
_{2} $CuO$_{4+\delta }$ from the elastic susceptibility and $^{139}$La NQR
relaxation rate versus temperature, as a low-frequency tail of the spectral
density of the atomic motions. The pseudo-diffusive motions have been
identified as the tilt modes of the O octahedra in a multi-well potential.
In the interpretation, we have used the fact that the dynamics of the
octahedra in the $ab$ plane can be cast into a non-linear one-dimensional
equation, which admits solitonic solutions corresponding to propagating
walls between domains with different tilt patterns. Thus it has been
possible to rely on the analysis of similar one-dimensional systems, for
which the spectral density contains a pseudo-diffusive central peak of width 
$\tau ^{-1}=\tau _{0}^{-1}\exp \left( -E_{\text{eff}}/k_{\text{B}}T\right) $
related to the soliton-like lattice excitations. The measured effective
barrier $E_{\text{eff}}/k_{\text{B}}=2800~$K is about 5 times higher than
the barrier of the theoretically estimated local potential, due to the
cooperative character of the motion. The alternating LTO domains and LTT
walls should correspond to the lattice stripes observed in superconducting
oxides.

\section{Figures captions}

Fig. 1. $^{139}$La NQR relaxation rates $W_{\text{Q}}$ (right hand scale)
and elastic energy loss $Q^{-1}$ (left hand scale) for a sample with $\delta
\sim 10^{-4}$, after subtraction of the contribution from the O diffusion.
Solid lines: best fits according to the model described in the text.

Fig. 2. Recovery plots of the amplitude of the free induction signal $M$
after saturation of the 3$\nu _{\text{Q}}$ NQR line (closed symbols and
solid lines) at 77 and 297$~$K. The open symbols are the recoveries
corresponding to the irradiation of the 2$\nu _{\text{Q}}$ NQR line. The
dashed and dotted lines are the curves expected in the case of pure
quadrupolar and pure magnetic relaxation mechanisms, respectively.

Fig. 3. Relaxation rate $\tau ^{-1}\left( T\right) $ corresponding to the
condition $\omega \tau =1$ at the maximum of the anelastic and NQR
relaxation curves (Fig. 1). The slope of the straight line provides the
effective activation energy $E_{\text{eff}}$, which coincides with the mean
activation energy of the best fit in Fig. 1.

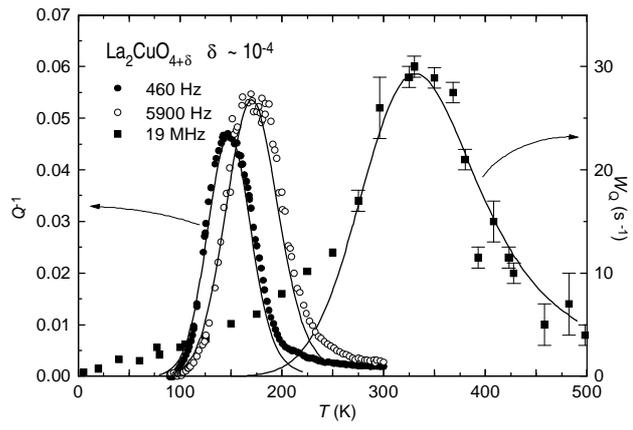
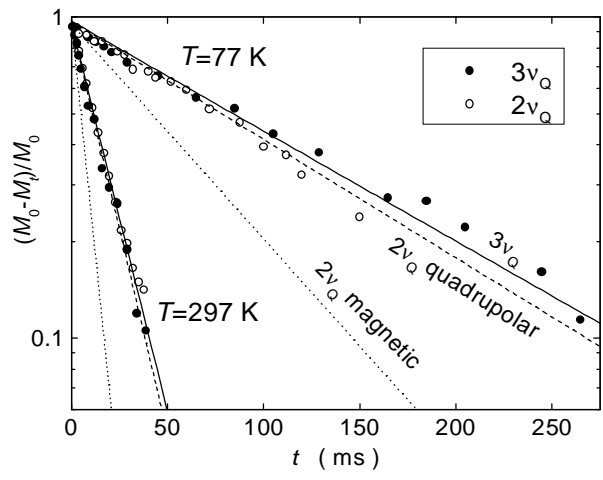
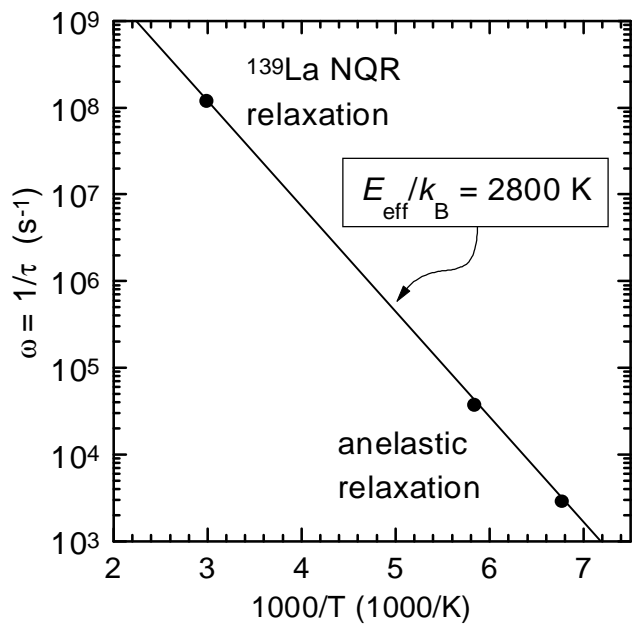
\end{document}